# Circular polarimetry reveals helical magnetic fields in the young stellar object HH135-136

Antonio Chrysostomou*, Phil W. Lucas*, & James H. Hough*

*Centre for Astrophysics Research, Science & Technology Research Institute, University of Hertfordshire, Hatfield, HERTS AL10 9AB, UK*

**Magnetic fields are believed to play a vital role in regulating and shaping the flow of material onto and away from protostars during their initial mass accretion phase. It is becoming increasingly accepted[1] that bipolar outflows are generated and collimated as material is driven along magnetic field lines and centrifugally accelerated off a rotating accretion disk. However, the precise role of the magnetic field is poorly understood and evidence for its shape and structure has not been forthcoming. Here we report, based on imaging circular polarimetry in the near-infrared and Monte Carlo modelling, that the magnetic field along the bipolar outflow of the HH135-136 young stellar object is helical. The magnetic field retains this shape for large distances along the outflow, giving *direct* evidence for a magnetic field structure which can also provide the necessary magnetic pressure for the outflow's collimation. Furthermore, this result lends further weight to the hypothesis, central to any theory of star formation, that the outflow is an important instrument for removal of high angular momentum material from the accretion disk thereby allowing the central protostar to increase its mass.**

Herbig-Haro (HH) objects were discovered as faint, nebulous emission on optical plates in the 1950's by G.H. Herbig[2] and G. Haro[3]. They possess characteristic optical spectra[4] and are regarded to be the result of a bipolar jet or outflow driven from a young stellar object (YSO) interacting with its parent molecular cloud[1]. In fact, the astrophysical jet is considered to be of major importance given its manifestation in a number of different environments from brown dwarfs[5] and binary systems[1] through to black holes and active galactic nucle[5]. Theoretical models which aim to explain the jet phenomenon now consistently appeal to the magneto-hydrodynamic (MHD) interaction of the magnetic field with the accreting gas in some form or another: X-wind models[6] generate an outflow very close to the star (~ few stellar radii) in the magnetopause between the accretion disk and the protostar; disk-wind models[6-7] generate an outflow over a relatively large range of radii of the disk surface. In each case, material is lifted and centrifugally accelerated along magnetic field lines (rather like 'beads on a wire'). Although present observational techniques struggle to resolve and probe into the innermost regions where the jet is launched, these theoretical models are able to explain some key observational properties, such as the correlation of accretion with ejection rates, and the fractional mass loss rates[7]. Observations are beginning to favour disk-wind models[8,9] and show that the outflow remains collimated over large distances, although the debate is far from resolved.

Mechanical collimation is possible if the ram pressure in the flow (acting normal to the cavity surface) is equal to or less than the pressure in the medium that the outflow is pushing through. Although such collimation is plausible for young sources which remain embedded within the natal molecular cloud, it does not explain the collimation seen in jets from optically revealed YSOs[10] – the so-called T-Taurii stars – which have had their circumstellar environment cleared away. In order to reconcile this, theories invoke the presence and action of magnetic fields whereby an initially collapsing cloud forms a rotating accretion disk in the plane perpendicular to the magnetic field lines which in turn twists the field lines into a helical field (with opposite helicity above and below the disk). This can provide the necessary magnetic pressure in the environment to retain the observed collimation to large distances via 'hoop' stresses[11]. Although evidence for strong (> 1 mG) magnetic fields at large distances from YSOs is available[12] it is small in number, while evidence for the morphology of this field is severely lacking.

The efficacy of polarimetry in constraining various physical parameters concerning the scattering geometry and media in the environments of young stellar objects has been demonstrated in numerous papers[13,14]. Grains are in general non-spherical and aligned by the ambient magnetic field, and selective absorption by the long axis of the grains polarizes the radiation (dichroism). Linear polarization (generated by dust scattering and/or dichroism) has an almost ubiquitous presence in star formation regions whereas circular polarization is far less common, requiring more specialized conditions for its generation, e.g. multiple scattering, scattering off aligned non-spherical grains (dichroic scattering) or dichroic extinction of linearly polarized light.

In Figure 1 we show near-infrared imaging circular polarimetry data for the HH135-136 outflow. Situated at a distance of 2.7 kpc in the Carina nebula[19] they are two of the most distant HH objects known, powered by an intermediate mass Herbig Ae-Be star, IRAS 11101-5829[20]. The data shows that the two lobes of the bipolar outflow are strongly circularly polarized. For the southern lobe, the bulk of the radiation is positively circularly polarized (i.e. right-handed in the direction of propagation) while negative circularly polarized light dominates the northern lobe. In each case, flips in the handedness of circular polarization are seen towards the limbs of the outflow lobes (most apparent in the $K_n$ and H band images). This pattern does not conform to the classical alternating symmetry seen in other objects and models[13-16], where the flip occurs along the axis of the flow and both senses of polarization are equally prominent. The degrees of circular polarization are relatively high. In the $K_n$ band the circular polarization is as much as $\sim -8\%$, becoming $\sim -3\%$ in the H band and $\sim -2.5\%$ in the J band. Values as high as $\sim$ 15-20% in the $K_n$ band have been reported towards the OMC-1 nebula in Orion[13] and in NGC6334V[21], both regions of high-mass star formation. The values we find here are significantly higher than those seen towards low-mass YSOs where absolute values < 1.5% are typically seen[14,22]. This places HH135-136 in an intermediate position between those objects exhibiting high values of circular polarization which are thus far associated with high-mass YSOs, and those with low values that are associated with low-mass YSOs. If one associates increasing magnetic field strength with higher mass star formation then this apparent correlation may be simply understood, as in the presence of a strong magnetic field the grains will align more efficiently and will more readily circularly polarize radiation.

It seems most likely that the only way to produce the pattern observed, while still retaining the observed centro-symmetric linear polarization pattern produced by scattering of light from the central protostar[20], is through the manipulation of the magnetic field structure. The classic symmetry of an alternating quadrupolar circular polarization pattern can arise from multiple scattering by spherical or non-aligned grains but this produces only ~1% polarization. If there are aligned non-spherical grains, the same pattern can be produced by dichroic scattering or through dichroic extinction if the field is oriented parallel to the outflow axis[15,17,23], and the fractional polarization is higher.

Output from a successful model (see *Supplementary Information*) with a pinched and twisted field structure is shown in Figure 2. We find that dichroic extinction (birefringence) is the principal cause of the observed 8% polarization. As well as reproducing the linear polarimetry[20], the model reproduces key features of the circular polarimetry: (*i*) the opposite sense of polarization in the two outflow lobes, which requires that the helicity reverses in the disk plane (i.e. a bihelical structure, which is naturally caused by a rotating disk); (*ii*) the flip in polarization at the eastern edge of the outflow in both lobes, at a certain distance along the flow, which is evidence that the pitch angle of the field to the disk plane increases with increasing distance (i.e. the field becomes more axial); (*iii*) the wavelength dependence of the polarization. This was contrary to the usual wavelength dependence of birefringence or dichroic scattering[18,23], but was found to be caused by the fairly high optical depth of the reflection nebulosity, which was determined from the near infrared colours ($A_V \approx 13$)[20]. At the shorter wavelengths the polarization is reduced by increased multiple scattering, since the optical depth and the grain albedo are higher at shorter wavelengths.

This optical depth implies a mass for the outflow of ~ 1 $M_\odot$ but given the high $H_2$ luminosity[24] and mass outflow rates[25] of objects typical of that driving HH135-136 (~ $10^{-3} - 10^{-4}$ $M_\odot$ yr$^{-1}$) this is not unfeasible. Optical and near-infrared images[19,24] show emission line structures coincident with the reflection nebulosity indicating that we are tracing the material, and hence magnetic field, in an outflow rather than an extended molecular envelope.

The field used is visualised in Figure 3. The model is not very sensitive to the radial component of the field (relative to the outflow axis) and so we cannot confirm whether the field lines are pinched in the disk plane, though the twist does appear to be strongest there. Observations at higher spatial resolution to more precisely measure the flips in polarisation at the eastern edge of the flow could in principle measure the field pitch angle as a function of distance along the flow, though the imperfect axisymmetry of the outflow would introduce some uncertainty.

We suggest that a helical field is a plausible model for the magnetic field structure about HH 135-136 (and other YSO outflows) and possesses an appropriate configuration to provide the necessary transverse pressure to retain the collimation of outflows, confirming an earlier suggestion based on linear polarimetry[26]. We also suggest that circular polarimetry represents a new and powerful technique for probing the magnetic field structure at these small scales (arcseconds on the sky) and calls for further observations to be made towards other YSOs. Finally, given that the combination of a rotating disk and a helical magnetic field structure would naturally launch material into the outflow with a significant toroidal component, angular momentum will be carried away from the central accreting system thus preventing centrifugal forces from stopping the collapse and allowing low angular momentum material to be accreted onto the protostar.

**Acknowledgements** The authors would like to thank M. Tamura for discussions on HH135 and for providing us with his linear polarimetry data to aid our analyses. This work was supported by a grant from the UK Science and Technologies Facilities Council (to P.W.L).

**Author Contributions** All authors contributed equally to this work. A.C and J.H.H conducted the observations at the telescope. A.C reduced the data while P.W.L carried out the Monte Carlo modelling. A.C wrote the main paper while P.W.L. wrote the Supplementary Information. All authors discussed the results and implications and commented on the manuscript at all stages.

**Author Information** The authors declare no competing financial interests. Correspondence and requests for materials should be addressed to A.C. (e-mail: a.chrysostomou@herts.ac.uk) or P.W.L (e-mail: p.w.lucas@herts.ac.uk).

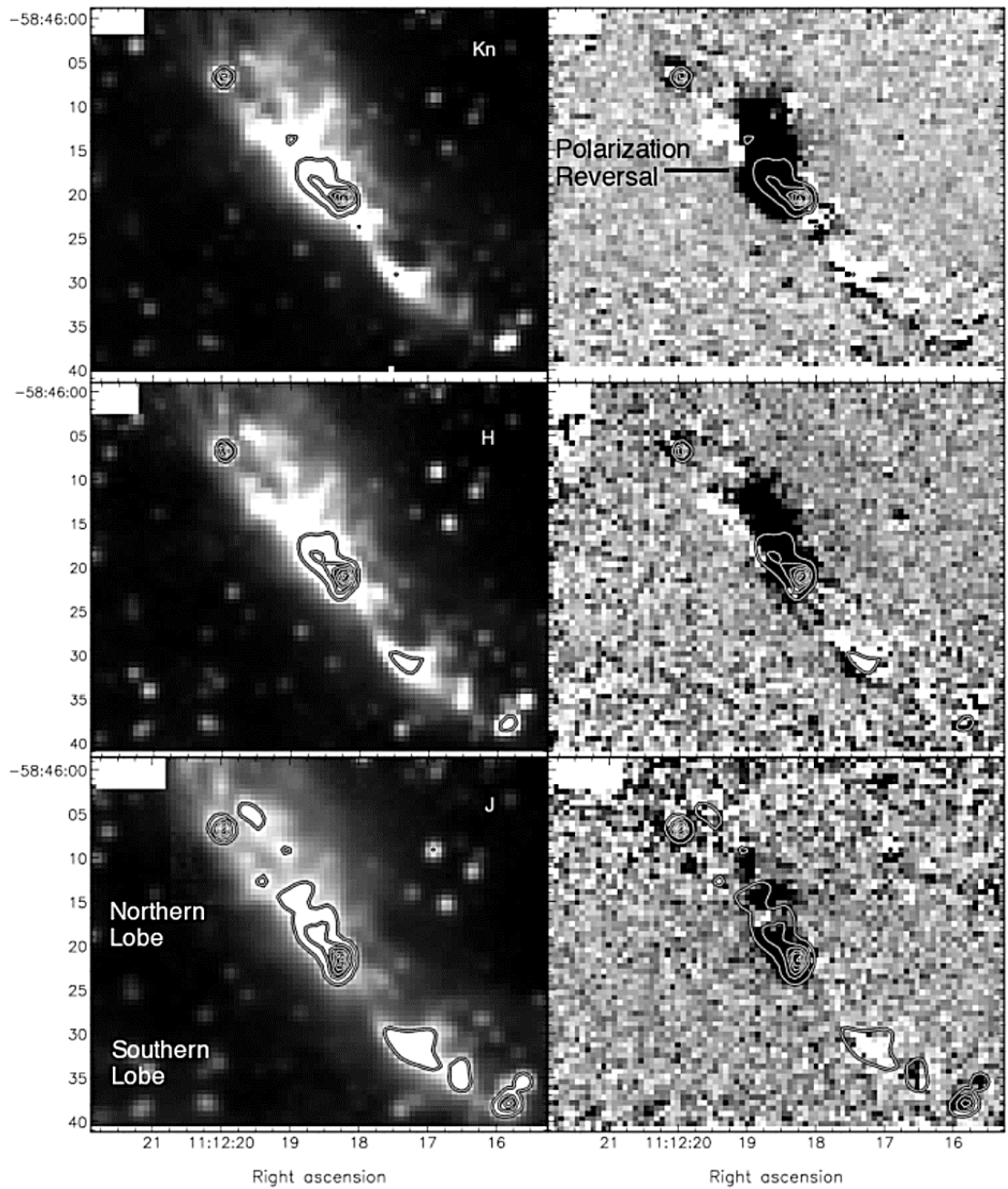

Figure 1: Imaging circular polarimetry of the HH135-136 outflow system. Polarimetry in the J, H and $K_n$ near-infrared atmospheric passbands was obtained at the Anglo-Australian Telescope on the nights of 22-24 May, 1997, using standard instrument and data reduction techniques[14]. The facility near-infrared imager-spectrometer (IRIS) was used at the *f*/15 telescope focus. The panels to the left show the near-infrared intensity in the $K_n$ (2.13 μm), H (1.6 μm) and J (1.25 μm) bands – top to bottom respectively. Coordinates are given

for epoch J2000, North is to the top and East to the left. The circular polarization (Stokes $V$) measured in these bands is shown in the right-hand panels. Intensity contours were arbitrarily chosen for each waveband, to accentuate the brighter emission features, and are plotted onto the intensity and circular polarization images. In each case, black is negatively circularly polarized and white is positive. By convention, positive polarization means that the electric vector is seen to rotate counterclockwise (right-handed) along the propagation direction in a fixed plane by an observer looking at the source. Note how each lobe is dominated by a single handedness of polarization and of opposite sign to the other lobe, although flips in the handedness of polarization are seen towards the edges of the lobes.

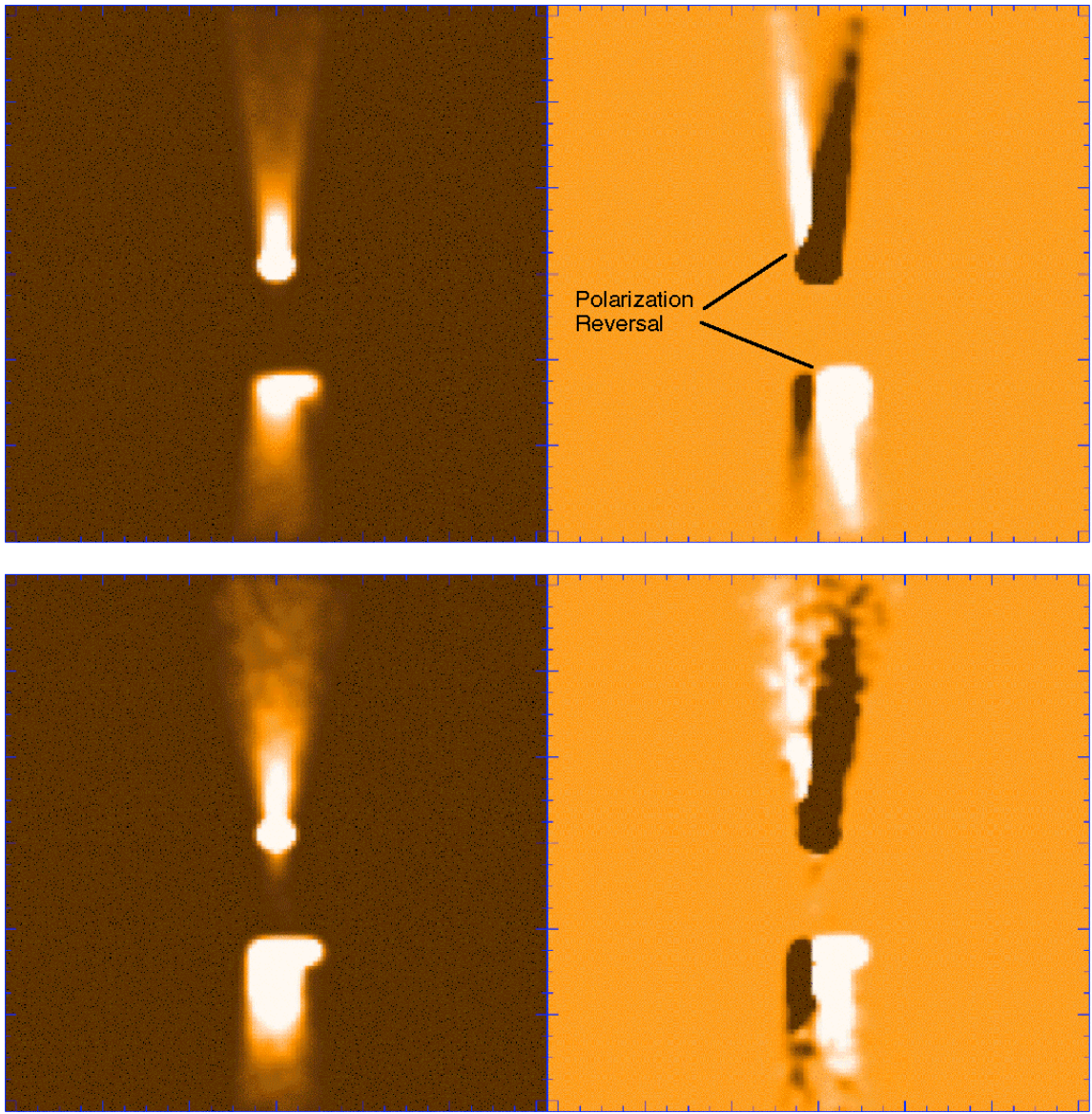

Figure 2: Results of 3-D Monte-Carlo light scattering model[17]. Intensity (left panel) and circular polarimetry (right panel) images are shown. The upper images are for the K band (2.2 µm wavelength) while the lower images are for the J band (1.25 µm wavelength). The light emanates from a central protostar and is scattered by dust in the bipolar outflow. The inner 50% of the outflow cross section (by radius) is optically thin, so that the full length of the outflow is illuminated. The outer 50% is denser and is responsible for producing the circular polarization through dichroic extinction. The protostar itself is entirely obscured from view by an optically thick accretion disk (500 AU radius) and a circumstellar envelope (1000 AU radius), which are viewed edge-on and therefore are also not seen. The bright peak near the centre of the model intensity images is due to scattering from the outflow in the upper lobe. The large vertical gap between the two outflow lobes is due to a much lower density of matter in the inner part of lower lobe. The model outflow has a non-axisymmetric structure in the lower lobe, in order to better reproduce the data. The structure of the model is fully described in the *Supplementary Information*.

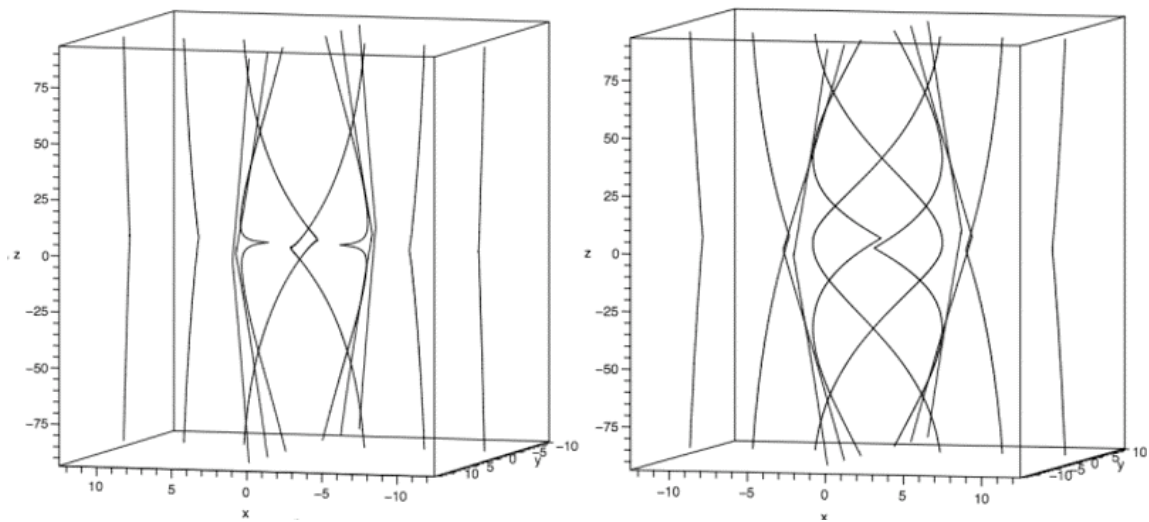

Figure 3. Helical magnetic field structures. The left panel shows the pinched and twisted magnetic field structure used in the model shown in Figure 2. The toroidal component, $B_\phi$, slowly decays away with distance from the source. This ensures that at large distances from the YSO, the magnetic field can remain contiguous with what is assumed to be a generally axial, large-scale, magnetic

field running through the molecular cloud. The right panel shows a field with no pinch in the disk plane (i.e. $B_r$=0) but with a slightly stronger twist. The two structures produce similar circular polarization. The field is drawn to scale with the data (Z = 100 = 27,000 AU). Z = 0 represents the disk plane. The sign of $B_z$ is the same in both hemispheres (i.e. dipole polarity). The sense of rotation of the disk is such that its rotation axis is parallel to North. The equations describing the field structure are given in the *Supplementary Information*.